%% file: paper.tex
\documentclass[11pt]{article}

\usepackage[a4paper,margin=2.5cm]{geometry}

\usepackage[auth-sc-lg,affil-sl]{authblk}
\usepackage{nomencl}
\usepackage{amsmath}
\usepackage{xcolor}
\usepackage{upgreek}

\include{format}

\newcommand{\rev}[1]{}

\usepackage{amssymb}
\usepackage{amsmath}
\usepackage[colorlinks=true, linkcolor=blue, citecolor=blue]{hyperref}
\usepackage{graphicx}
\usepackage{nomencl}
\usepackage{threeparttable}
\usepackage[marginal,bottom]{footmisc}

\usepackage{tikz}
\usepgflibrary{arrows}

\title{A Micromechanics-Based Model for Stiffness and Strength Estimation of
Cocciopesto Mortars}

\author{V\'{a}clav Ne\v{z}erka}
\author{Jan Zeman}
\affil{Department of Mechanics, Faculty of Civil Engineering,
  Czech Technical University in Prague, Th\'{a}kurova 7, 166 29 Prague
  6, Czech Republic}
  
\begin{document}
\maketitle

\begin{abstract}%
The purpose of this paper is to propose an inexpensive micromechanics-based
scheme for stiffness homogenization and strength estimation of mortars
containing crushed bricks, known as cocciopesto. The model utilizes the
Mori-Tanaka method to determine the effective stiffness, combined with estimates
of quadratic invariants of the deviatoric stresses inside phases to predict the
compressive strength. Special attention is paid to the representation of C-S-H
gel layer around bricks and interfacial transition zone around sand aggregates,
which renders the predictions sensitive to particle sizes. Several parametric
studies are performed to demonstrate that the method correctly reproduces data
and trends reported in available literature. Moreover, the model is based
exclusively on parameters with clear physical or geometrical meaning and as such
it provides a convenient framework for its further experimental validation.
\end{abstract}


\section{Introduction}
The use of lime as a binder in mortars is associated with well-known
inconveniences such as slow setting and carbonation, high drying shrinkage and
porosity and low mechanical strength~\cite{Arizzi_2012}. Although these
limitations have been overcome with the use of Portland cement in the last
$50$~years, lime mortars still find use in the restoration of historic
structures. This is mainly due to their superior compatibility with the original
materials in contrast to many modern renovation render systems,
e.g.~\cite{Maravelaki_2003,Michoniova_2005,Sepulcre_2010}.

Mechanical properties of lime mortars can be improved by a suitable
design of the mixture. Phoenicians were probably the first ones who added
crushed clay products, such as burnt bricks, tiles or pieces of pottery, to lime
mortars to increase their durability and strength. Romans called such material
\emph{cocciopesto} and utilized this mortar in areas where other natural
pozzolans were not available. The cocciopesto-based structures exhibit
increased ductility, leading to their remarkable resistance to
earthquakes~\cite{Baronio_1997,Moropoulou:2002:ABC}.

Much later, it was found that the mortars containing crushed clay bricks, burnt
at $600$--$900^\circ$C, exhibit a hydraulic character, manifested by the
formation of a thin layer of Calcium-Silicate-Hydrate~(C-S-H) gel at the
lime-brick interface~\cite{Moropoulou_1995}. Since C-S-H gel is the key
component responsible for favorable mechanical performance of Portland cement
pastes~\cite{Neville_1996}, it is generally conjectured that the enhanced
performance of cocciopesto mortars can be attributed to the high strength and
stiffness of the C-S-H gel
coating~\cite{Baronio_1997,Moropoulou_1995,Moropoulou:2002:ABC,Sepulcre_2010}.
This mechanism competes with the formation of the Interfacial Transition
Zone~(ITZ) at the matrix-aggregate interface, known to possess higher porosity
and thus lower stiffness in cement-based mortars,
e.g.~\cite{Ollivier:1995:ITZ,Scrivener:2004:ITZ,Yang_1998}.

The purpose of this work is to interpret these experimental findings by a
micromechanical model based on the Mori-Tanaka method~\cite{Mori_1973},
motivated by its recent applications to related material systems. These include,
for example, estimates of effective thermal conductivity of rubber-reinforced
cement composites~\cite{Stransky:2011:MTB}, elasticity predictions for early-age
cement~\cite{Bernard:2003:MMH} or alkali-activated~\cite{Smilauer_2011} pastes,
upscaling the compressive strength of cement mortars~\cite{Pichler_2011}, and
multi-scale simulations of three-point bending tests of concrete
specimens~\cite{Vorel_2012}. Here, we exploit these developments to propose a
simple analytical model for stiffness and strength estimation of cocciopesto
mortars in \Sref{sec:model}. In particular, the elasticity predictions utilize
Benveniste's reformulation~\cite{Benveniste_1987} of the Mori-Tanaka
method~\cite{Mori_1973}, whereas the strength predictions build on recent
results by Pichler and Hellmich~\cite{Pichler_2011}, who demonstrated that
compressive strength is closely related to the quadratic average of the
deviatoric stress in the weakest phase. A particular attention is paid to
representation of coatings by C-S-H gel and ITZ, which renders the
predictions sensitive to the size of brick particles and aggregates. In
\Sref{sec:Results}, we verify predictions of the proposed scheme against data
available in open literature. These finding are summarized in
\Sref{sec:Conclusions}, mainly as a support for future validation of the model
against experimental results. Finally, in \Aref{app:Herve-Zaoui} we gather
technical details needed to account for coated inclusions in order to make the
paper self-contained.

In what follows, the Mandel representation of symmetric tensorial quantities is
systematically employed, e.g.~\cite[p.~23]{Milton:2002:TC}. In particular,
italic letters, e.g. $a$ or $A$, refer to scalar quantities and  boldface
letters, e.g. $\M{a}$ or $\M{A}$, denote vectors or matrix representations of
second- or fourth-order tensors. $\M{A}\trn$ and $(\MA)^{-1}$ standardly denote
the matrix transpose and the inverse matrix. Other symbols and abbreviations are
introduced in the text when needed.

\section{Model}\label{sec:model}

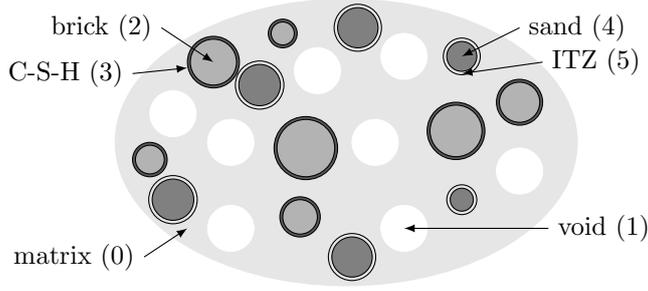
\begin{figure}
\centering%
\begin{tikzpicture}[scale=.76]
\filldraw[black!10](0,0) ellipse(4 and 2.5); 
\filldraw[white](1,1.5) circle (0.4);%
\filldraw[white](-3,.5) circle (0.4); %
\filldraw[white](-.5,1.25) circle (0.4);%
\filldraw[white](0.5,0) circle (0.4); %
\filldraw[white](1,-1.5) circle (0.4);%
\filldraw[white](3,-.5) circle (0.4); %
\filldraw[white](-2,0) circle (0.4); %
\filldraw[white](-2,-1.5) circle (0.4); %
\filldraw[black!10, draw=black](2,1.5) circle (0.32);%
\filldraw[black!50, draw=black](2,1.5) circle (0.26);%
\filldraw[black!10, draw=black](-1.5,1.) circle (0.42); %
\filldraw[black!50, draw=black](-1.5,1.) circle (0.36); %
\filldraw[black!10, draw=black](0.2,2) circle (0.41); %
\filldraw[black!50, draw=black](0.2,2) circle (0.35); %
\filldraw[black!10, draw=black](2,-1.) circle (0.26); %
\filldraw[black!50, draw=black](2,-1.) circle (0.20); %
\filldraw[black!10, draw=black](-3,-1.0) circle (0.42); %
\filldraw[black!50, draw=black](-3,-1.0) circle (0.36); %
\filldraw[black!10, draw=black](.1,-2.) circle (0.41); %
\filldraw[black!50, draw=black](.1,-2.) circle (0.35); %
\filldraw[black!65, draw=black](3,0.7) circle (0.4);%
\filldraw[black!30, draw=black](3,0.7) circle (0.35);%
\filldraw[black!65, draw=black](-2.3,1.4) circle (0.45);%
\filldraw[black!30, draw=black](-2.3,1.4) circle (0.40);%
\filldraw[black!65, draw=black](-0.8,-1.3) circle (0.35);%
\filldraw[black!30, draw=black](-0.8,-1.3) circle (0.3); %
\filldraw[black!65, draw=black](-3.4,-0.3) circle (0.30);
\filldraw[black!30, draw=black](-3.4,-0.3) circle (0.25);
\filldraw[black!65, draw=black](-1.1,1.9) circle (0.25);%
\filldraw[black!30, draw=black](-1.1,1.9) circle (0.20);%
\filldraw[black!65, draw=black](1.9,0.2) circle (0.5);%
\filldraw[black!30, draw=black](1.9,0.2) circle (0.45);%
\filldraw[black!65, draw=black](-0.7,-0.1) circle (0.55);%
\filldraw[black!30, draw=black](-0.7,-0.1) circle (0.50);%
%
\small
\coordinate[label=left:{matrix (0)}] (1) at (-3.5,-2);
\draw[latex-] (-2.75,-1.5) -- (1);%
\coordinate[label=left:{brick (2)}] (2) at (-3.2,2);
\draw[latex-] (-2.3,1.4) -- (2);%
\coordinate[label=right:{void (1)}] (2) at (3.5,-1.5);
\draw[latex-] (1.0,-1.5) -- (2);%
\coordinate[label=right:{sand (4)}] (3) at (3,2);
\draw[latex-] (2.0,1.5) -- (3);%
\coordinate[label=right:{ITZ (5)}] (4) at (3.4,1.4);
\draw[latex-] (2.0,1.2) -- (4);%
\coordinate[label=left:{C-S-H (3)}] (5) at (-3.7,1.2);
\draw[latex-] (-2.75,1.3) -- (5);%
\end{tikzpicture}
\caption{Scheme of the micromechanics-based model. The numbers in parentheses
refer to indexes of individual phases.}
\label{fig:scheme}
\end{figure}

We consider a composite sample occupying domain $\rve$, composed of $\nphs$
distinct phases indexed by $\iphs$. The value $r=0$ is reserved for the matrix
phase and $\iphs = 1, \ldots, \nphs$ refer to heterogeneities having the shape
of a sphere or spherical shell, see~\Fref{fig:scheme}. Volume fraction of the
$\iphs$-th phase are defined as $\vfrac\phs{\iphs} = |\rve\phs{\iphs}| /
|\rve|$, where $|\rve\phs{\iphs}|$ denotes the volume occupied by the $\iphs$-th
phase, and geometry of coated particles is specified by their radii
$\rd\phs{\iphs}$ for $\iphs =2, \ldots, 5$, \Fref{fig:coated}.
\nomenclature{$\rve$}{Representative volume element RVE}%
\nomenclature{$\rve\phs{\iphs}$}{Part of RVE occupied by $\iphs$-th phase}%
\nomenclature{$\iphs$}{Index used to number phases}%
\nomenclature{$\vfrac$}{Volume fraction}%

\rev{%
Several comments are in now order concerning simplifications adopted in the
model. First, brick particles and voids are considered to be of the spherical
shape, instead of more realistic ellipsoids as in
e.g.~\cite{Pichler:2009:SAR,Pichler_2011}. This step is known to introduce only
minor errors in the prediction of overall transport~\cite{Stransky:2011:MTB} or
elastic~\cite{Pichler:2009:SAR} properties. As demonstrated by Pichler et
al.~\cite{Pichler:2009:SAR}, the up-scaled strength is more sensitive to the
shape of inhomogeneities, but the model is still capable to predict the correct
trends. Second, the ITZ is taken as homogeneous and is not resolved down to the
level of microheterogeneities. This arises as a result of the fact that, in
contrast to cement-based materials~\cite{Ollivier:1995:ITZ,Scrivener:2004:ITZ},
we are currently not aware of any work studying the structure of ITZ in
lime-based mortars, thus input data for a more detailed representation are
missing. Third, only monodisperse distribution of particles in assumed.
Polydispersivity can be incorporated by simple averaging arguments and results
only in a moderate increase of accuracy~\cite[and references
therein]{Stransky:2011:MTB}.}

Elastic properties of individual  phases are specified by the material
stiffness matrix $\ML\phs{\iphs}$. As each phase is assumed to be homogeneous
and isotropic, we have
\nomenclature{$\ML\phs{\iphs}$}{Stiffness matrix of the $\iphs$-th phase}%
\begin{equation}\label{eq:phase_stiffness}
\ML\phs{\iphs} 
=
3\K\phs{\iphs}
\MIV
+
2\G\phs{\iphs}
\MID
\quad \text{ for }
\iphs = 0, \ldots, \nphs,
\end{equation}
where $\K\phs{\iphs}$ and $\G\phs{\iphs}$ are the bulk and shear moduli of the
$\iphs$-th phase, and $\MIV$ and $\MID$ denote the orthogonal projections to the
volumetric and deviatoric components, e.g.~\cite{Jirasek:2007:BCE}, so that
\begin{subequations}\label{eq:mean_stress_strain}
\begin{align}
\Meps(\x)
& =
( \MIV + \MID )\Meps(\x) 
= 
\epsV(\x) \M{1} + \MepsD(\x),
\\
\Msig(\x)
& =
( \MIV + \MID ) \Msig(\x) 
=
\sigV(\x) \M{1} + \MsigD(\x),
\end{align}
\end{subequations}
for $\x \in \rve$. In~\Eqref{eq:mean_stress_strain}, $\Meps$ and $\Msig$ refer
to local stresses and strains, $\epsV$ and $\MepsD$ are the volumetric and
deviatoric strain components, $\sigV$ and $\MsigD$ refer to the stress
components and $\M{1}$ is the second-order unit tensor (in the matrix
representation).

Development of the model follows the standard routine of the continuum
micromechanics, e.g.~\cite{Zaoui:2002:CMS}. The sample $\rve$ is subjected to
the overall strain loading $\MEps$. Neglecting the interaction among phases, the
mean strains inside heterogeneities are obtained as
\nomenclature{$\MEps$}{Average strain}%
\begin{equation*}
\MEps\phs{\iphs}
=
\MA\dil\phs{\iphs}
\MEps
\quad \text{ for }
\iphs = 1, \ldots, \nphs,
\end{equation*}
where $\MA\dil\phs{\iphs}$ is the dilute concentration factor of the $\iphs$-th
phase, see \Sref{sec:dilute}. In \Sref{sec:stiffness}, after accounting for the
phase interaction, these are combined to the full concentration factors
satisfying
\begin{equation*}
\MEps\phs{\iphs}
=
\MA\phs{\iphs}
\MEps
\quad \text{ for }
\iphs = 0, \ldots, \nphs,
\end{equation*}
\nomenclature{$\MA\phs{\iphs}$}{Strain concentration factor of the $\iphs$-th
phase}%
utilized next to estimate the overall stiffness of the composite material,
$\ML\eff$. Moreover, as outlined in~\Sref{sec:strength}, expression for the
overall stiffness also encodes the mean value of the quadratic invariant of the
local stress deviator $\MsigD$ defined as
\begin{equation}\label{eq:J2_def}
\J\phs{\iphs} 
=
\sqrt{%
\frac{1}{2 |\rve\phs{\iphs}|}
\int_{\rve\phs{\iphs}}
\MsigD(\x)\trn \MsigD(\x)
\de\x},
\end{equation}
\nomenclature{$\J$}{Second invariant of the stress deviator}%
\nomenclature{$\MsigD$}{Local stress deviators}%
\nomenclature{$\x$}{Position vector}%
that can be directly used to estimate the overall strength of a material.

\subsection{Dilute concentration factors}\label{sec:dilute}
Due to geometrical and material isotropy of individual phases, the dilute
concentration factors attain the form analogous to~\eqref{eq:phase_stiffness}:
\begin{equation}
\MA\dil\phs{\iphs}
=
\AdilK{\iphs} \MIV + \AdilG{\iphs} \MID
\text{ for }
r = 1, \ldots, n.
\end{equation}
The expressions for the components are given separately for the
\emph{uncoated}~($r=1$) and \emph{coated}~($r=2, \ldots, 5$) particles.
Namely, in the first case it holds
\begin{align*}
\AdilK{1} 
& = 
\frac{%
\K\phs{0}
}{
\K\phs{0} 
+ 
\alpha\phs{0} 
(\K\phs{1} - \K\phs{0})
},
\\
\AdilG{1} 
& = 
\frac{%
\G\phs{0}
}{
\G\phs{0} 
+ 
\beta\phs{0} 
(\G\phs{1} - \G\phs{0})
},
\end{align*}
with the auxiliary factors following from the Eshelby
solution~\cite{Eshelby_1957} in the form 
\begin{align*}
\alpha\phs{0} 
= 
\frac{1+\nu\phs{0}}{3(1+\nu\phs{0})},
&&
\beta\phs{0} 
= 
\frac{2(4-5\nu\phs{0})}{15(1-\nu\phs{0})},
\end{align*}
where $\nu\phs{0}$ is the Poisson ratio of the matrix phase.

\begin{figure}
\centering
\begin{tikzpicture}[scale=.75]
\filldraw[black!10](-2.5,-2.5) rectangle(2.5,2.5); 
\filldraw[black!20, draw=black](0,0) circle (1.75);%
\filldraw[black!50, draw=black](0,0) circle (1.5);%
\draw[arrows=->, line width=0.4pt](0,0)--(-1.4,-1.05);
\draw[arrows=->, line width=0.4pt](0,0)--(0.6,1.375);
\small
\node[fill=white, inner sep=1pt] at (-2,-2) {$(i_3)$};
\node[fill=white, inner sep=1pt] at (-0.7,0.4) {$(i_1)$};
\node[fill=white, inner sep=1pt] at (1.2,-1.2) {$(i_2)$};
\coordinate[label=left:{$\rd\phs{i_1}$}] (1) at (1.4,0.5);
\coordinate[label=right:{$\rd\phs{i_2}$}] (1) at (-0.8,-0.65);
\end{tikzpicture}
\caption{Scheme of a single-layer inclusion}%
\label{fig:coated}
\end{figure}
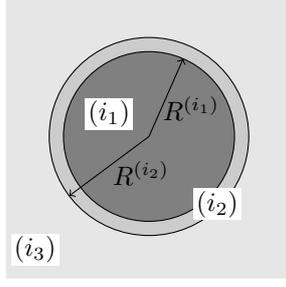

The coated case is more involved, and was first solved in its full generality 
by Herve and Zaoui~\cite{Herve_1993} for multi-layered spherical inclusion. To
apply their results in the current setting, we locally number the phases by the
index $\M{i} = [ i_1, i_2, i_3 ]\trn$, see \Fref{fig:coated}, where $\M{i} =
[2,3,0]\trn$ for the brick--C-S-H conglomerate and $\M{i} = [4,5,0]\trn$ refers
to a sand particle coated by ITZ. Now, we have
\nomenclature{$\M{i}$}{Index used for coated inclusions}%
\begin{align}\label{eq:layered_volumetric}
\AdilK{i_1} = \frac{1}{Q_{11}^2}, 
&& 
\AdilK{i_2} = \frac{Q_{11}^1}{Q_{11}^2},
\end{align}
and
\begin{align}\label{eq:layered_deviatoric}
\AdilG{i_1}
& = 
A_1-\frac{21}{5}
\frac{\rd\phs{i_1}{}^2}{1-2\nu\phs{i_1}}
B_1,
\\
\AdilG{i_2}
& = 
A_2
-
\frac{21}{5}
\frac{%
\rd^{(i_2)}{}^5
-
\rd^{(i_2-1)}{}^5
}{
(1-2\nu^{(i_2)})
(\rd^{(i_2)}{}^3
-\rd^{(i_2-1)}{}^3)
}
B_2,
\nonumber
\end{align}
where the auxiliary factors are provided in \Aref{app:Herve-Zaoui}.

\begin{table*}
 \centering
 \begin{threeparttable}
 \caption{Reference properties of individual phases; $\rho$ denotes \rev{mass}
 density, $f_\mathrm{t}$ is tensile strength, $m$ is the mass fraction\rev{, and
 radii $\rd$ defined according to \Fref{fig:coated}}.}
 \label{tab:properties}
\begin{tabular}{clccccccl}
  \hline
  $\iphs$ & Phase & $\rho$ & $E$ & $\nu$ & $f_\mathrm{t}$ & $m$ & $\rd$
  & Note \\
  & & [kgm$^{-3}$] & [MPa] & [-] & [MPa] & [-] & [$\upmu$m] & \\
  \hline
  $0$ & Pure lime matrix & 
  $1,200$\tnote{a} & 
  $2,000$\tnote{\cite{Drdacky_2003}} & 
  $0.25$\tnote{\cite{Drdacky_2003}} & 
  $0.4$\tnote{\cite{Drdacky_2003}} & 
  $3$ & $\times$ & \\
  $1$ & Voids & $\times$ & $10^{-9}$ & $0.25$ & $\times$ & 
  $\times$ & \rev{$0.1$--$100$}\tnote{\cite{Mosquera_2006}} & 
  \hspace{2mm}$\vfrac\phs{1} = 35\%$\tnote{a} \\
  $2$ & Clay brick & 
  $2,300$\tnote{a} & 
  $5,000$\tnote{a} & 
  $0.17$\tnote{a} & 
  $3.2$\tnote{a} & 
  $1$ & $500$ \\
  $3$ & C-S-H gel & 
  $2,000$\tnote{\cite{Jeffrey_2006}} & 
  $22,000$\tnote{\cite{Constantinides_2004}} & 
  $0.2$\tnote{\cite{Constantinides_2004}} & 
  $\times$ & 
  $\times$ & $510$\tnote{\cite{Boke_2006}} & \\
  $4$ & Siliceous sand & 
  $2,700$\tnote{a} & 
  $60,000$\tnote{\cite{Vorel_2012}} & 
  $0.17$\tnote{\cite{Vorel_2012}} & 
  $48$\tnote{\cite{Neville_1996}} &
  $1$ & $500$ & \\
  $5$ & ITZ & 
  $1,200$\tnote{b} & 
  $500$\tnote{c} & 
  $0.25$\tnote{b} & 
  $\times$ &
  $\times$ & $520$\tnote{c} &  \\
  \hline
\end{tabular}
\begin{tablenotes}
\item[a] Our own (unpublished) data. \rev{Densities and porosity were
measured by a pycnometer, elastic constants were determined from strain-gauge
data in compression test, and tensile strength follows from a unidirectional
tensile test.}
\item[b] Same value as for the lime matrix.
\item[c] Set as in~\cite{Yang_1998} for cement-based concretes,
i.e. Young's modulus to $20$--$40\%$ of the value for the matrix phase and
thickness to $20~\upmu$m.
\end{tablenotes}
\end{threeparttable}
\end{table*}

\subsection{Stiffness estimates}\label{sec:stiffness}
In Benveniste's~\cite{Benveniste_1987} interpretation of the original
Mori-Tanaka method~\cite{Mori_1973}, the mutual interaction among
heterogeneities is modeled by loading each particle by the average strain in
the matrix phase $\MEps\phs{0}$ instead of $\MEps$. For this purpose, we relate
$\MEps\phs{0}$ to $\MEps$ by a strain compatibility condition, valid under the
dilute approximation,
\begin{equation*}
\MEps 
= 
\bigl( 
\vfrac\phs{0} \MI
+
\sum_{\iphs=1}^{\nphs}
\vfrac\phs{\iphs} 
\MA\dil\phs{\iphs}
\bigr)
\MEps\phs{0},
\end{equation*}
from which we express the full concentration factors as
\begin{align*}
\MA\phs{0}
& =
\bigl( 
\vfrac\phs{0} \MI
+
\sum_{\iphs=1}^{\nphs}
\vfrac\phs{\iphs} 
\MA\dil\phs{\iphs}
\bigr)^{-1},
\\
\MA\phs{\iphs}
& = 
\MA\dil\phs{\iphs} \MA\phs{0}
\text{ for }
r = 1, \ldots, \nphs.
\end{align*}
Utilizing a universal relation
\begin{equation*}
\ML\eff
=
\sum_{\iphs=0}^{\nphs}
\vfrac\phs{\iphs}
\ML\phs{\iphs}
\MA\phs{\iphs},
\end{equation*}
we can see that the effective stiffness inherits the symmetry of individual
phases~\eqref{eq:phase_stiffness} with 
\begin{subequations}\label{eq:effective_constants}
\begin{align}
\K\eff 
& = 
\frac{%
\vfrac\phs{0} 
\K\phs{0} 
+ 
\sum_{\iphs=1}^{\nphs}
\vfrac\phs{\iphs}
\K\phs{\iphs} \AdilK{\iphs} 
}{%
\vfrac\phs{0} 
+
\sum_{\iphs=1}^{\nphs}
\vfrac\phs{\iphs} \AdilK{\iphs}
}, \\
\G\eff 
& =
\frac{%
\vfrac\phs{0} 
\G\phs{0} 
+ 
\sum_{\iphs=1}^{\nphs}
\vfrac\phs{\iphs}
\G\phs{\iphs} \AdilG{\iphs} 
}{%
\vfrac\phs{0} 
+
\sum_{\iphs=1}^{\nphs}
\vfrac\phs{\iphs} \AdilG{\iphs}
}.
\end{align}
\end{subequations}

\subsection{Strength estimates}\label{sec:strength}

As recognized first by Kreher~\cite{Kreher:1990:RSS}, the fluctuations of
stresses and strains in individual phases can be estimated from the energy
conservation condition due \rev{to Hill~\cite{Hill:1963:EPRS}:}
\begin{equation}\label{eq:energy_equality}
\begin{split}
\MEps\trn 
\ML\eff 
\MEps
= &
\frac{1}{|\rve|}
\sum_{\iphs=0}^{\nphs}
\Bigl(
9\K\phs{r}
\int_{\rve\phs{r}}
\epsV^2(\x)
\de\x
\\
& + 
\int_{\rve\phs{r}}
2\G\phs{r}
\MepsD\trn(\x)
\MepsD(\x)
\de \x
\Bigr),
\end{split}
\end{equation}
\nomenclature{$\epsV$}{Volumetric part of strain}%
\nomenclature{$\MepsD$}{Deviatoric part of strain}%
expressing the conservation of energy on macroscale due to $\MEps$ and the
average local values due to $\epsV$ and $\MepsD$.
Differentiating~\eqref{eq:energy_equality} with respect to $\G\phs{r}$, we
obtain
\begin{equation*}
\MEps\trn 
\frac{\partial \ML\eff}{\partial \G\phs{r}} 
\MEps
=
\frac{2}{|\rve|}
\int_{\rve\phs{r}}
\MepsD\trn(\x)
\MepsD(\x)
\de \x,
\end{equation*}
for $\iphs = 0, \ldots, \nphs$. Next, we recognize that $\MsigD\phs{r} =
(1/2\G\phs{r}) \MepsD\phs{r}$ and recall the definition of the quadratic
invariant~\eqref{eq:J2_def} to arrive at
\begin{equation}\label{eq:J2_comp}
\J\phs{\iphs}
=
\G\phs{\iphs}
\sqrt{%
\frac{1}{\vfrac\phs{\iphs}}
\MEps\trn 
\frac{\partial \ML\eff}{\partial \G\phs{r}} 
\MEps
}.
\end{equation}

As thoroughly demonstrated by \rev{Pichler et al.~\cite{Pichler:2009:SAR}} and
Pichler and Hellmich~\cite{Pichler_2011}, this quantity is closely related to
the compressive strength $\fc$ of cement pastes at various degrees of hydration.
Here, we postulate that
\nomenclature{$\fc$}{Compressive strength}%
\begin{equation}\label{eq:strength_scaling}
\frac{\fc(\p_1)}{\fc(\p_2)}
\approx
\frac{\J\phs{\w}(\p_2)}{\J\phs{\w}(\p_1)},
\end{equation}
where $\w = 0, \ldots, \nphs$ is the index of the weakest phase and~$p$ refers
to a parameter characterizing the mixture composition, see the next section for
concrete examples.
\nomenclature{$\p$}{Composition parameter}%
\nomenclature{$\w$}{Index of the weakest phase}%

\section{Results and discussion}\label{sec:Results}
The purpose of this section is to examine the trends in mechanical properties as
predicted by the proposed scheme. Default data for individual phases, summarized
in \Tref{tab:properties}, were partly assembled from open literature and
complemented with our own, yet unpublished, measurements. Note that the
matrix--brick--sand fractions correspond to a typical composition of historic
lime mortars~\cite{Baronio_1997,Baronio_1997_2} and that the engineering
constants $E$ and $\nu$ are connected to the bulk and shear moduli through
well-known relations, e.g.~\cite[p.~23]{Milton:2002:TC},
\begin{align*}
K
=
\frac{E}{3(1 - 2\nu)},
&& 
G 
=
\frac{E}{2(1 + \nu)}. 
\end{align*}

\rev{%
Given the data in \Tref{tab:properties}, the volume fraction of individual
phases are determined from six independent conditions. The first two relate the volume
fractions of brick and sand particles and their coatings by
\begin{align*}
\vfrac\phs{3} 
= 
\Bigl( \bigl( 
\frac{\rd\phs{3}}{\rd\phs{2}} 
\bigr)^3 - 1 
\Bigr)
\vfrac\phs{2}
, &&
\vfrac\phs{5} 
= 
\Bigl( \bigl( 
\frac{\rd\phs{5}}{\rd\phs{4}} 
\bigr)^3 - 1 
\Bigr)
\vfrac\phs{4}.
\end{align*}
Next, we enforce the value of mass fractions via
\begin{align*}
\vfrac{\phs{0}}
=
\frac{m\phs{0}\rho\phs{2}}{m\phs{2}\rho\phs{0}}
\vfrac{\phs{2}},
&&
\vfrac{\phs{0}}
=
\frac{m\phs{0}\rho\phs{4}}{m\phs{4}\rho\phs{0}}
\vfrac{\phs{4}},
\end{align*}
where $m\phs{\iphs}$ and $\rho\phs{\iphs}$ denote the mass fraction and the
mass density of the $\iphs$-th phase, respectively. Since $\vfrac\phs{1}$ is
given, the remaining condition is provided by
$
\sum_{\iphs=0}^{5}
\vfrac\phs{\iphs}
= 1
$, and the phase volume fractions follow as the solution of the 
system of $6\times 6$ linear equations.

In the sensitivity analyzes}, motivated by experimental findings in
e.g.~\cite{Baronio_1997,Velosa_2009,Vejmelkova_2012}, we assume that an increase
in the C-S-H gel volume fraction~($\Delta\vfrac\phs{3}$) is compensated by the
corresponding changes for  matrix~($\Delta\vfrac\phs{0}$),
voids~($\Delta\vfrac\phs{1}$), and clay bricks~($\Delta\vfrac\phs{2}$), so that
\begin{equation*}
\Delta\vfrac\phs{0} + \Delta\vfrac\phs{1} + \Delta\vfrac\phs{2} +
\Delta\vfrac\phs{3} = 0,
\end{equation*}
where we set for simplicity $\Delta\vfrac\phs{1} = \Delta\vfrac\phs{2} =
\Delta\vfrac\phs{3}$. Analogously, an increase in the volume fraction of ITZ
corresponds to the decrease in volume of the matrix phase:
\begin{equation*}
\Delta\vfrac\phs{0} + \Delta\vfrac\phs{5} = 0.
\end{equation*}

In the strength estimates, the imposed loading simulates the
uniaxial compression test, for which $\MSig = [-1, 0, 0, 0, 0, 0 ]\trn$ and the
average strain follows from
\begin{equation*}
\MEps = (\ML\eff)^{-1} \MSig.
\end{equation*}
\nomenclature{$\Msig$}{Average stress}%
We assume that ITZ is the weakest phase, i.e. $w = 5$ in
\Eqref{eq:strength_scaling}, and\rev{, similarly to~\cite{Pichler:2009:SAR},}
estimate the derivative in \Eqref{eq:J2_comp} by the forward difference with the step size of $\Delta
G\phs{5} = 1$~Pa.\footnote{%
Our results are reproducible with a MATLAB code {\bf Homogenizator~MT}, freely
available at
\url{http://mech.fsv.cvut.cz/~nezerka/software}.}

\begin{figure}
\centering
\begin{tabular}{ll}
 \includegraphics[width=.48\textwidth]{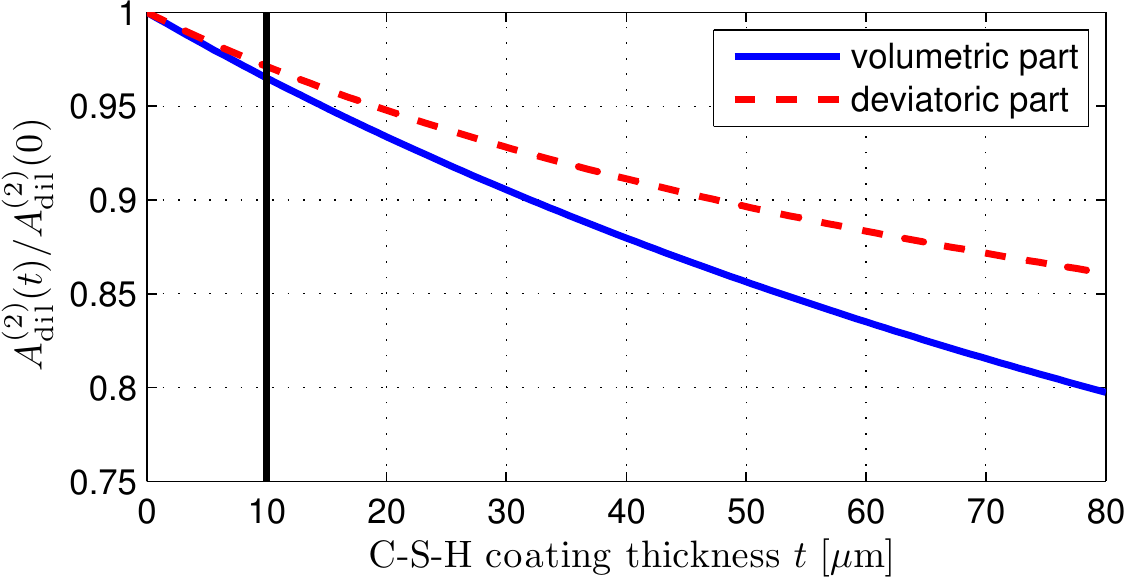} &
 \includegraphics[width=.48\textwidth]{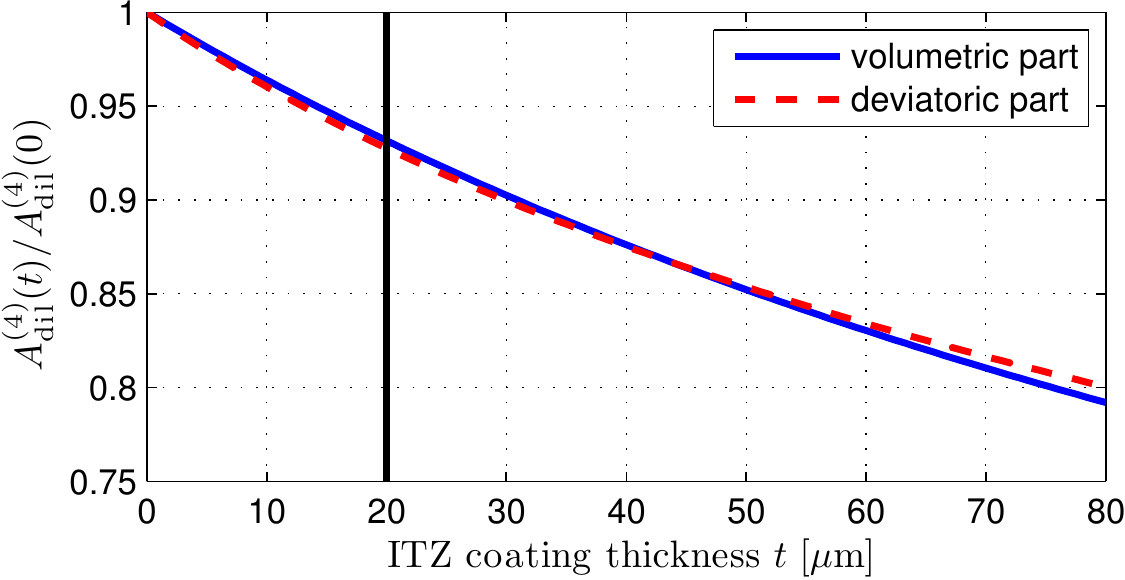} \\[-5mm]
 (a) & (b) \\[-2mm]
 \end{tabular} 
 \caption{Influence of the coating thickness on the dilute
 concentration factors for (a)~brick and (b)~sand particles. \rev{The vertical
 lines refer to the default thicknesses that are kept constant in the remaining
 sensitivity analyzes.}}
 \label{fig:coating}
\bigskip
\centering
\begin{tabular}{ll}
 \includegraphics[width=.48\textwidth]{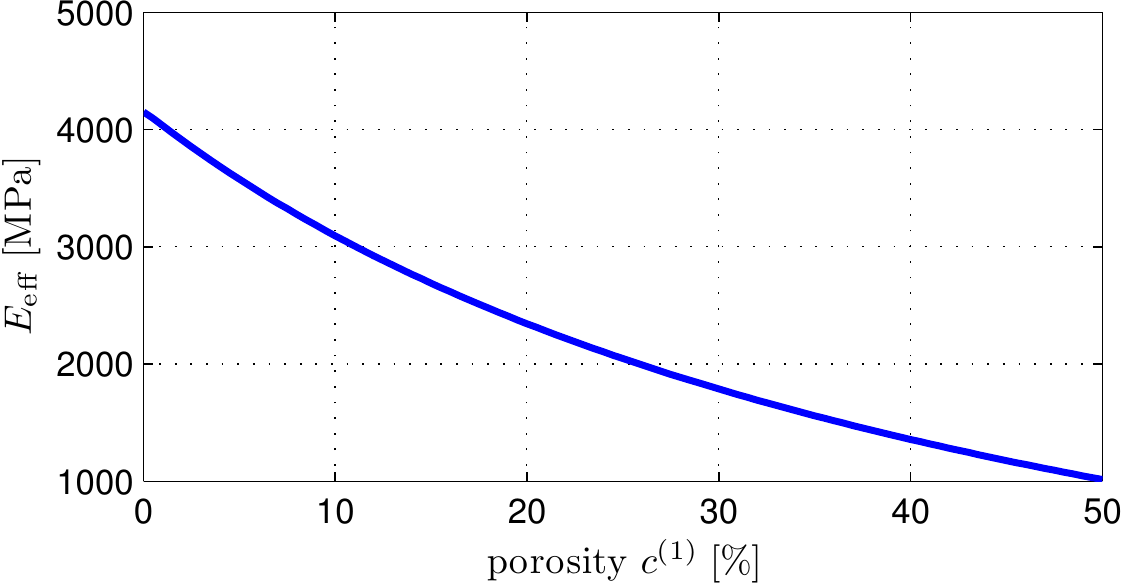} &
 \includegraphics[width=.48\textwidth]{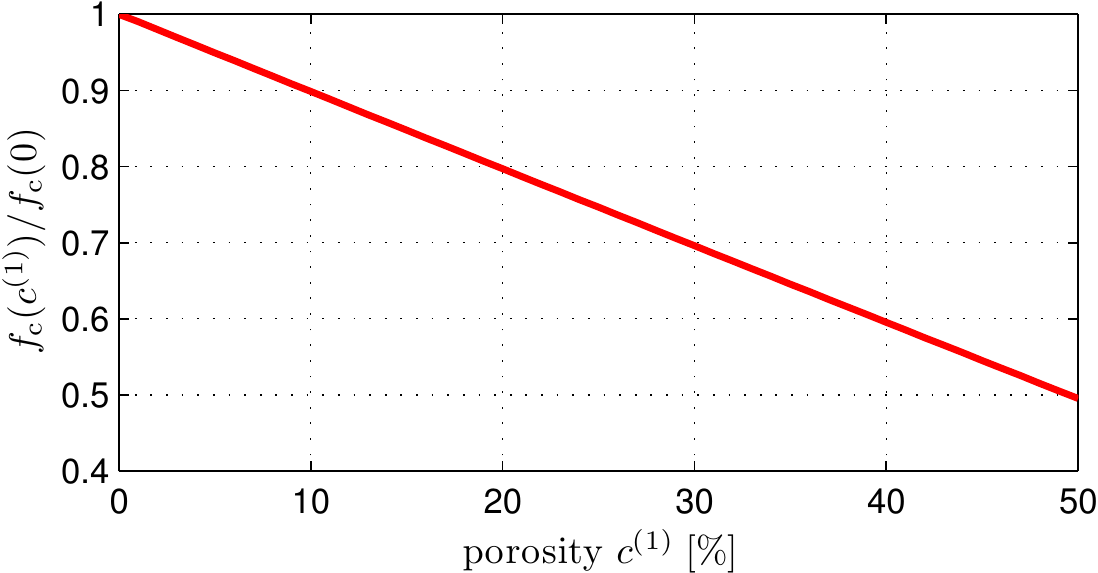} \\[-5mm]
 (a) & (b) \\[-2mm]
\end{tabular}
 \caption{(a)~Stiffness- and (b)~strength-porosity relations.}
 \label{fig:porosity}
\end{figure}

\subsection{Effect of coatings}

The first aspect we would like to discuss is the effect of coating on the
dilute concentration factors of the brick and sand particles. \Fref{fig:coating}
demonstrates that, in terms of the volumetric phase strains, the effects of
C-S-H and ITZ are comparable, despite the fact the C-S-H is stiffer and ITZ
more complaint than the matrix phase. The differences in the deviatoric part,
which drive the strength estimates according to \Eqref{eq:strength_scaling},
become more pronounced with the increasing thickness. This indicates that the
contribution of brick and sand particles to the overall properties might still
be different, once accounting phase properties and their interaction, 
see \Sref{sec:size_effects} for further discussion.

\subsection{Influence of porosity}\label{sec:porosity}

Analogously to the cement pastes, porosity has a major influence on the overall
properties of lime-based mortars. This is also confirmed by the results of the
proposed model shown in \Fref{fig:porosity}. As for the overall stiffness, for
the realistic range of porosities of $25$--$40 \%$~\cite{Drdacky_2003}, the
estimates~\eqref{eq:effective_constants} predict the values of Young modulus
between $\approx 2,000$ and $1,000$~MPa. This is consistent with the values
reported in~\cite{Baronio_1997} historic lime mortars~(without pozzolan
admixtures).

As for the strength estimates, it follows from \Fref{fig:porosity}~(b) that they
reproduce \rev{a power-law relation}~\cite[p.~280]{Neville_1996}
\begin{equation}\label{eq:Powers}
\fc(\vfrac\phs{1})
= 
\fc(0)
( 1 - \vfrac\phs{1})^n,
\end{equation}
with $n \doteq 1.04$, yielding practically the linear strength-porosity scaling.
Unfortunately, we are currently unable to validate this prediction against
experiments; the only available work we are aware of by Papayianni and
Stefanidou~\cite{Papayianni_2005} does not contain enough data. Still,
\Eqref{eq:Powers} complies with the fact the influence of porosity is much
smaller in lime mortars than in cement-based
materials~\rev{\cite[Section~2.6]{Lawrence:2006:SCN}}, for which $n \approx 3$
is typically used, \rev{see \cite[and references therein]{Papayianni_2005}}.

\subsection{Size effects}\label{sec:size_effects}
\begin{figure}
 \begin{tabular}{ll}
 \includegraphics[width=.48\textwidth]{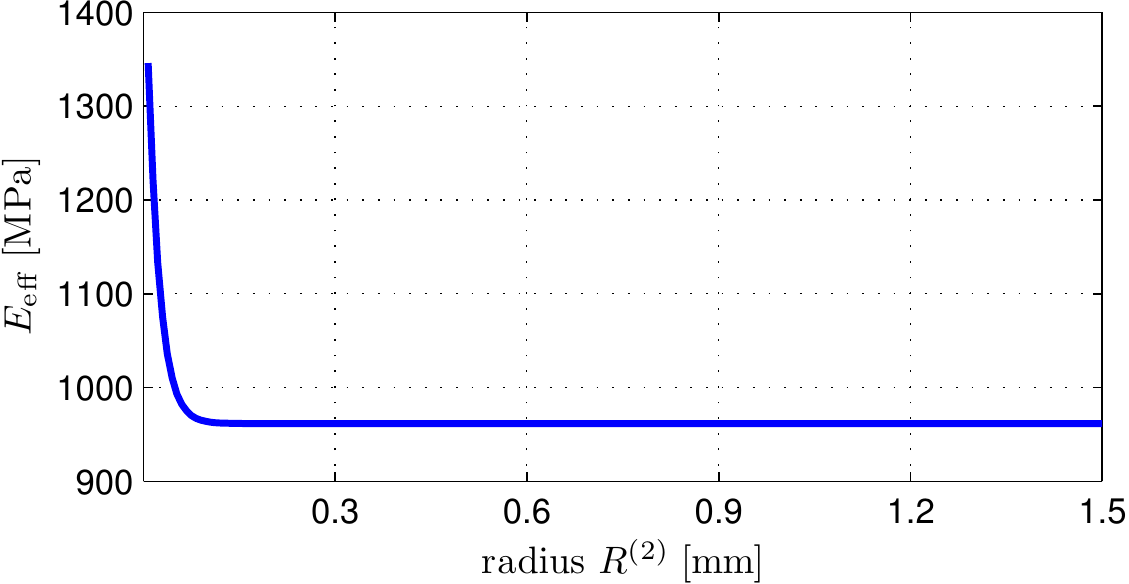} &
 \includegraphics[width=.48\textwidth]{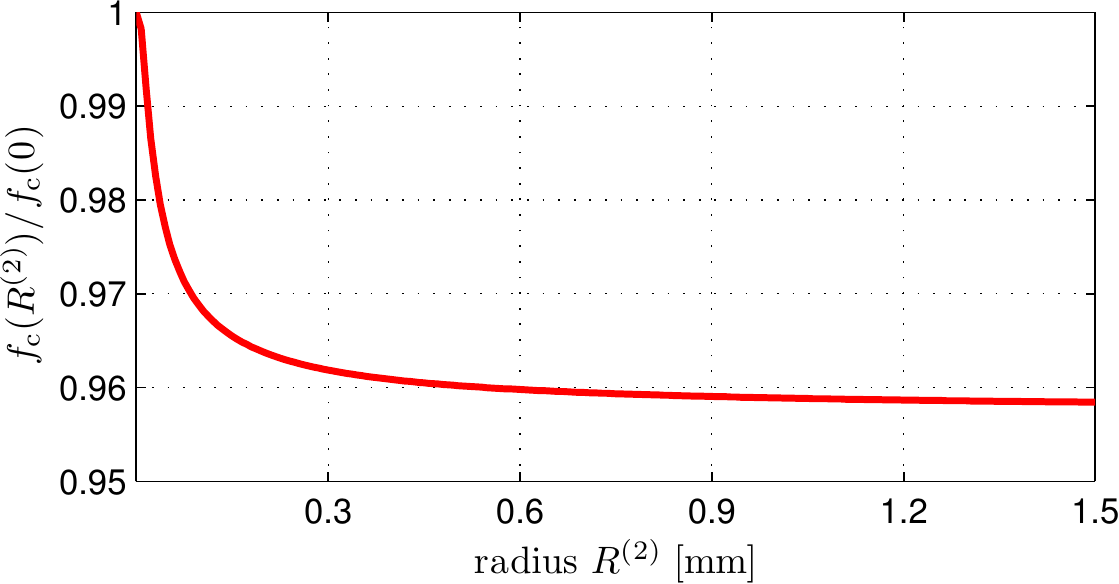} \\[-5mm]
 (a) & (b) \\[-2mm]
 \end{tabular} 
 \caption{Influence of the brick particle size on the overall (a)~stiffness and
 (b)~strength.}
 \label{fig:size_effect_brick}
\bigskip
 \begin{tabular}{ll}
 \includegraphics[width=.48\textwidth]{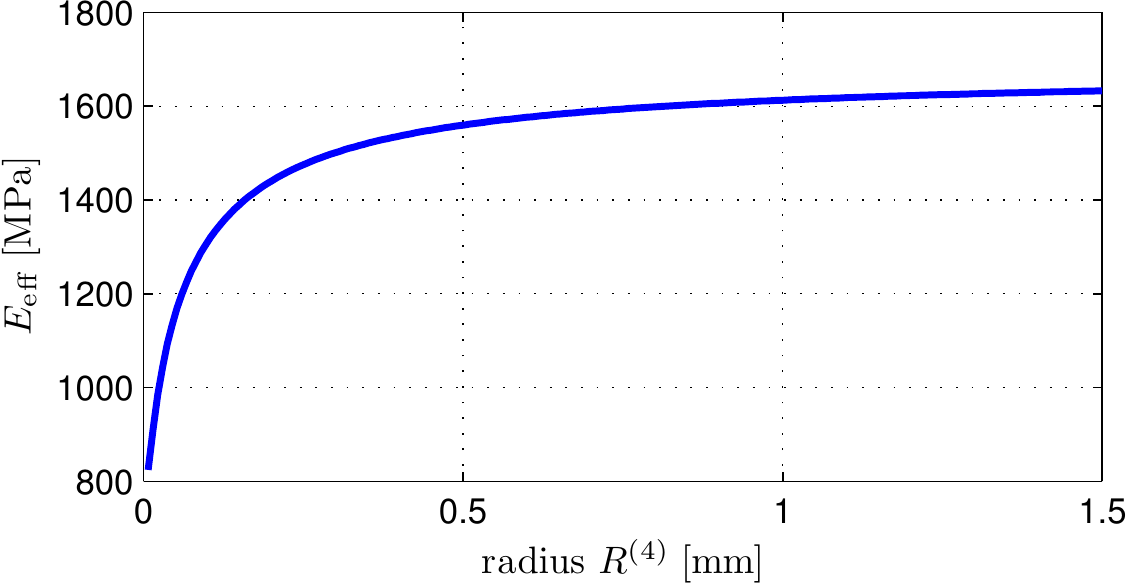} &
 \includegraphics[width=.48\textwidth]{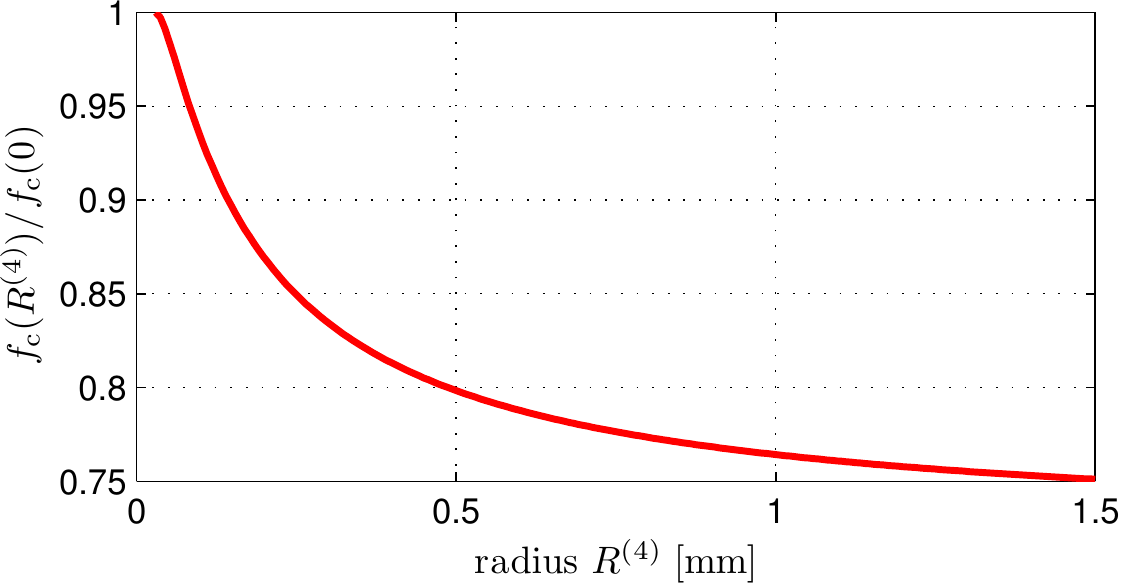} \\[-5mm]
 (a) & (b) \\[-2mm]
 \end{tabular} 
 \caption{Influence of the sand particle size on the overall (a)~stiffness and
 (b)~strength.}
 \label{fig:size_effect_sand}
\end{figure}

Now we proceed to clarify the impact of brick and sand
particles on the overall mechanical properties.\footnote{%
\rev{In the sensitivity analysis, the mass ratio is kept constant according to
\Tref{tab:properties}.}}
In particular, when
increasing the size of brick particles, the material becomes more complaint since the
stiffening effect of C-S-H layer decreases, \Fref{fig:size_effect_brick}(a).
This also increases the deviatoric stresses in ITZ, as manifested by the
strength reduction visible in \Fref{fig:size_effect_brick}(b). These effects
practically stabilize for particles larger than $0.5$~mm and their magnitude is
rather limited: the stiffness decreases by about $10~\%$ and the strength only
by $4~\%$. Such trends are qualitatively consistent with the results presented
e.g. in~\cite{Baronio_1997,Moropoulou:2002:ABC,Sepulcre_2010}.

Larger sand particles, on the other hand, tend to make the composite material
stiffer, \Fref{fig:size_effect_sand}(a), by compensating for inferior mechanical
properties of ITZ. Since the relative thickness of ITZ layer decreases, the
stresses inside this phase increase and the material becomes weaker in overall,
\Fref{fig:size_effect_sand}(b). When compared to brick particles, these effects
are much more pronounced: in the considered range of radii, the Young modulus
increases by about $100~\%$ and the strength decreases by $25~\%$ with no
tendency to stabilize. This agrees well with experimental outcomes reported
in~\cite{Stefanidou_2005}.

\section{Conclusions}\label{sec:Conclusions}
In the present work, following the recent developments presented
in~\cite{Pichler:2009:SAR,Pichler_2011,Smilauer_2011,Vorel_2012}, a simple
micromechanics-based scheme for strength and stiffness estimates of cocciopesto
mortars has been presented. The model utilizes directly measurable material and
geometrical properties of individual phases and is free of adjustable
parameters. On the basis of presented results, we conclude that the model

\begin{enumerate}\itemsep=0pt
  \item predicts realistic values of the overall Young modulus and the
  strength-porosity scaling,
  \item captures the ``smaller is stiffer'' and ``smaller is
  stronger'' trends for crushed brick particles,
  \item captures the ``larger is stiffer'' and ``larger is weaker'' trends for
  sand aggregates,
  \item explains the positive role of crushed bricks in comparison with
  sand aggregates.
\end{enumerate}

Of course, in order to accept this model for practical use, it needs to be
validated against comprehensive experimental data at micro- and macro-scales,
\rev{and the role of ITZ in lime-based mortars needs to be clarified.} This
topic is currently under investigation and will be reported separately.

\paragraph{Acknowledgments}
\rev{We wish to thank an anonymous referee for a number of valuable
suggestions on the previous version of the manuscript.} This work was supported
by the Grant Agency of the Czech Technical University in Prague, project
No.~SGS12/027/OHK1/1T/11 (VN, JZ) and by the Ministry of Education, Youth and
Sports of the Czech Republic, project No.~684077003~(JZ).

\appendix

\section{Herve-Zaoui solution}\label{app:Herve-Zaoui}

The effect of coating on the mechanical properties enters the solution through
the auxiliary factors $\M{Q}^k$ in \Eqref{eq:layered_volumetric}, and $\M{A}^k$
and $\M{B}^k$ in \Eqref{eq:layered_deviatoric}. Here, these are provided in the
closed form optimized for coding, utilizing the results and nomenclature by
Herve and Zaoui~\cite{Herve_1993}. Note that in order to keep the notation
consistent, $a\phs{k}$ corresponds to a property of the $k$-th phase, whereas
$a^k$ denotes a quantity utilized in the Herve-Zaoui solution~(independent of
$a\phs{k}$). Also recall that we employ the local numbering of phases by the
index $\M{i} = [ i_1, i_2, i_3 ]\trn$ introduced by \Fref{fig:coated}.

In particular, the volumetric part is expressed in terms of matrices 
\begin{align*}
\M{Q}^1 = \M{N}^1, &&
\M{Q}^2 = \M{N}^2 \M{Q}^1,
\end{align*}
with
\begin{equation*}
\M{N}^k
=
\frac{1}{3K\phs{i_{k+1}}+G\phs{i_{k+1}}}
\begin{bmatrix}
3K\phs{i_k} + 4G\phs{i_{k+1}} & 
\dfrac{4}{\rd\phs{i_k}{}^3}(G\phs{i_{k+1}}-\G\phs{i_k}) \\
3\rd\phs{i_k}{}^3(K\phs{i_{k+1}}-K\phs{i_k}) &
3K\phs{i_{k+1}} + 4G\phs{i_k} 
\end{bmatrix}
\text{ for }
k = 1, 2.
\end{equation*}

The matrices needed to evaluate the deviatoric part follow from
\begin{equation*}
\M{A}^1 
= 
\frac{%
P_{22}^2}{
P_{11}^2 P_{22}^2 -P_{12}^2 P_{21}^2 },
\quad
\M{A}^2
=
W_1^2,
\quad
\M{B}^1
=
\frac{%
-P_{21}^2}{
P_{11}^2 P_{22}^2 -P_{12}^2 P_{21}^2 },
\quad
\M{B}^2 = W_2^2,
\end{equation*}
where
\begin{equation*}
\M{W}^k
= 
\frac{1}{P_{22}^2P_{11}^2-P_{12}^2P_{21}^2}
\M{P}^{k-1} 
\begin{bmatrix}
 P_{22}^2 &
-P_{21}^2 &
 0 &
 0
\end{bmatrix}\trn
(k = 1, 2),
\quad
\M{P}^1 = \M{M}^1,
\quad
\M{P}^2 = \M{M}^2 \M{P}^1.
\end{equation*}
The auxiliary matrix $\M{M}^k$ admits the expression:
\begin{equation*}
\M{M}^k
=
\frac{1}{5(1-\nu\phs{i_{k+1}})}
\begin{bmatrix}
\dfrac{c^k}{3} & 
\dfrac{\rd\phs{i_k}{}^2(3b^k-7c^k)}{5(1-2\nu\phs{i_k})} & 
\dfrac{-12\alpha^k}{\rd\phs{i_k}{}^5} & 
\dfrac{4(f^k-27\alpha^k)}{15\rd\phs{i_k}{}^3(1-2\nu\phs{i_k})} \\
0 & 
\dfrac{b^k(1-2\nu\phs{i_{k+1}})}{5(1-2\nu\phs{i_k})} &
M_{23}^k &
\dfrac{-12\alpha^k(1-2\nu\phs{i_{k+1}})}{7\rd\phs{i_k}{}^7(1-2\nu\phs{i_k})} \\
\dfrac{\rd\phs{i_k}{}^5\alpha^k}{2} &
\dfrac{-\rd\phs{i_k}{}^7(2a^k+147\alpha^k)}{70(1-2\nu\phs{i_k})} &
\dfrac{d^k}{7} & M_{34}^k
 \\
M_{41}^k &
\dfrac{7\alpha^k\rd\phs{i_k}{}^5(1-2\nu\phs{i_{k+1}})}{2(1-2\nu\phs{i_k})} &
0 & 
\dfrac{e^k(1-2\nu\phs{i_{k+1}})}{3(1-2\nu\phs{i_k})} 
\end{bmatrix}
\end{equation*}
with
\begin{align*}
M_{23}^k
& =
\dfrac{-20\alpha^k(1-2\nu\phs{i_{k+1}})}{7\rd\phs{i_k}{}^7}, 
&
M_{41}^k
=
\dfrac{-5\alpha^k\rd\phs{i_k}{}^3(1-2\nu\phs{i_{k+1}})}{6}, \\
M_{34}^k
& =
\frac{\rd\phs{i_k}{}^2(105(1-\nu\phs{i_{k+1}})
+
12\alpha^k(7-10\nu\phs{i_{k+1}})-7e^k)}{35(1-2\nu\phs{i_k})},
\end{align*}
and
\begin{align*}
a^k &= \frac{G\phs{i_k}}{G\phs{i_{k+1}}}
(7+5G\phs{i_k})(7-10G\phs{i_{k+1}})-(7-10G\phs{i_k})(7+5G\phs{i_{k+1}}), \\ 
b^k &= \frac{G\phs{i_k}}{G\phs{i_{k+1}}} (7+5G\phs{i_k})+4(7-10G\phs{i_k}), \\
c^k &= (7-5G\phs{i_{k+1}})+ 2(4-5G\phs{i_{k+1}}) \frac{G\phs{i_k}}{G\phs{i_{k+1}}}, \\ 
d^k &= (7+5G\phs{i_{k+1}})+ 4(7-10G\phs{i_{k+1}}) \frac{G\phs{i_k}}{G\phs{i_{k+1}}},\\ 
e^k &= 2(4-5G\phs{i_k})+ \frac{G\phs{i_k}}{G\phs{i_{k+1}}} (7-5G\phs{i_k}), \\ 
f^k &= (4-5G\phs{i_k})(7-5G\phs{i_{k+1}})- \frac{G\phs{i_k}}{G\phs{i_{k+1}}}
(4-5G\phs{i_{k+1}})(7-5G\phs{i_k}), \\ 
\alpha^k &= \frac{G\phs{i_k}}{G\phs{i_{k+1}}}-1.
\end{align*}

\end{document}

%% file: format.tex
\newcommand{\rve}{\Omega}
\newcommand{\nphs}{n}
\newcommand{\iphs}{r}
\newcommand{\phs}[1]{^{(#1)}}
\newcommand{\M}[1]{{\boldsymbol #1}}

\newcommand{\Sref}[1]{Section~\ref{#1}}
\newcommand{\Aref}[1]{Appendix~\ref{#1}}
\newcommand{\Fref}[1]{Fig.~\ref{#1}}
\newcommand{\Eqref}[1]{Eq.~\eqref{#1}}
\newcommand{\Tref}[1]{Table~\ref{#1}}

\newcommand{\vfrac}{c}
\newcommand{\ML}{\M{L}}
\newcommand{\rd}{R}

\newcommand{\K}{K}
\newcommand{\G}{G}
\newcommand{\MIV}{\M{I}_\mathrm{V}}
\newcommand{\MID}{\M{I}_\mathrm{D}}

\newcommand{\Meps}{\M{\varepsilon}}

\newcommand{\MEps}{\M{E}}%
\newcommand{\MSig}{\M{\Sigma}}%

\newcommand{\Msig}{\M{\sigma}}
\newcommand{\sigV}{\sigma_\mathrm{V}}

\newcommand{\J}{J_2}%
\newcommand{\MsigD}{\Msig_\mathrm{D}}%
\newcommand{\x}{\M{x}}%
\newcommand{\trn}{^{\sf T}}
\newcommand{\de}{\,{\mathrm d}} 

\newcommand{\MA}{\M{A}}
\newcommand{\dil}{_\mathrm{dil}}
\newcommand{\eff}{_\mathrm{eff}}
\newcommand{\MI}{\M{I}}

\newcommand{\AdilK}[1]{A_\mathrm{dil,V}\phs{#1}}
\newcommand{\AdilG}[1]{A_\mathrm{dil,D}\phs{#1}}

\newcommand{\epsV}{\varepsilon_\mathrm{V}}%
\newcommand{\MepsD}{\M{\varepsilon}_\mathrm{D}}%

\newcommand{\fc}{f_\mathrm{c}}
\newcommand{\p}{p}
\newcommand{\w}{w}